\documentclass[journal]{IEEEtran}
\IEEEoverridecommandlockouts

\usepackage[utf8]{inputenc}
\usepackage{amsmath}
\usepackage{amssymb}
\usepackage{amsthm}
\usepackage{enumerate}
\usepackage{graphics} 
\usepackage{graphicx}

\usepackage{algorithm}
\usepackage[noend]{algpseudocode}

\usepackage{tikz}
\usetikzlibrary{arrows,shapes,chains,matrix,positioning,scopes,patterns,calc,
decorations.markings,
decorations.pathmorphing,
}

\usepackage{pgfplots}
\pgfplotsset{compat=1.3}
\usepgflibrary{shapes}

\newcommand{\ev}{\ensuremath{\mathbf{e}}}

\newcommand{\rv}{\ensuremath{\mathbf{r}}}
\newcommand{\sv}{\ensuremath{\mathbf{s}}}

\newcommand{\vv}{\ensuremath{\mathbf{v}}}
\newcommand{\wv}{\ensuremath{\mathbf{w}}}
\newcommand{\xv}{\ensuremath{\mathbf{x}}}
\newcommand{\yv}{\ensuremath{\mathbf{y}}}
\newcommand{\zv}{\ensuremath{\mathbf{z}}}

\newcommand{\zerov}{\ensuremath{\boldsymbol{0}}}
\newcommand{\alphav}{\ensuremath{\boldsymbol{\alpha}}}
\newcommand{\etav}{\ensuremath{\boldsymbol{\eta}}}
\newcommand{\zetav}{\ensuremath{\boldsymbol{\zeta}}}

\newcommand{\Am}{\ensuremath{\mathbf{A}}}

\newcommand{\Dm}{\ensuremath{\mathbf{D}}}

\newcommand{\IDm}{\ensuremath{\mathbf{I}}}

\newcommand{\Phim}{\ensuremath{\mathbf{\Phi}}}

\newcommand{\Ncal}{\ensuremath{\mathcal{N}}}
\newcommand{\Ocal}{\ensuremath{\mathcal{O}}}

\newcommand{\Scal}{\ensuremath{\mathcal{S}}}

\newcommand{\hfrak}{\ensuremath{\mathfrak{h}}}
\newcommand{\rfrak}{\ensuremath{\mathfrak{r}}}
\newcommand{\sfrak}{\ensuremath{\mathfrak{s}}}

\title{Multi-User SR-LDPC Codes}
\author{Jamison R. Ebert, \textit{Student Member, IEEE}, Jean-Francois Chamberland, \textit{Senior Member, IEEE}, \\Krishna R. Narayanan, \textit{Fellow, IEEE}
\thanks{
This material is based upon work supported, in part, by the National Science Foundation (NSF) under Grants CCF-2131106 \& CNS-2148354.
This work was presented, in part, at the 2024 IEEE International Symposium on Information Theory (ISIT). 

Jamison R. Ebert, Jean-Francois Chamberland, and Krishna R. Narayanan are with the Department of Electrical and\ Computer Engineering, Texas A\&M University, College Station, TX 77843, USA (emails: \{jrebert, chmbrlnd, krn\}@tamu.edu).
\textit{Corresponding author: J.-F. Chamberland.}
}
}

\begin{document}

\maketitle

\begin{abstract}
This article introduces a novel non-orthogonal multiple access (NOMA) scheme for coordinated uplink channels.
The scheme builds on the recently proposed sparse-regression low-density parity-check (SR-LDPC) code, and extends the underlying notions to scenarios with many concurrent users.
The resulting scheme, called Multi-User SR-LDPC (MU-SR-LDPC) coding, consists of each user transmitting its own SR-LDPC codeword using a unique sensing matrix in conjunction with a characteristic outer LDPC code.
To recover the sent information, the decoder jointly processes the received signals using a low-complexity and highly-parallelizable AMP-BP algorithm.
At finite blocklengths (FBL), MU-SR-LDPC codes are shown to achieve a target BER at a higher spectral efficiency than baseline orthogonal multiple access (OMA) and non-orthogonal multiple access (NOMA) schemes with similar computational complexity. 
Furthermore, MU-SR-LDPC codes are shown to match the performance of maximum a posteriori (MAP) decoding in certain regimes.
For certain blocklengths and a sufficiently high number of users, MU-SR-LDPC codes may achieve a higher spectral efficiency than the approximate FBL capacity of the effective single-user Gaussian channel seen by each user in a comparable OMA scheme. 
Results are supported by numerical simulations.
\end{abstract}


\section{Introduction}
\label{section:introduction}

Coordinated multiple access schemes have been a hallmark of wireless communication systems for decades, enabling a multitude of users to share network resources in a fair and efficient manner.
Historically, orthogonal multiple access (OMA) schemes have been preferred in wireless standards, in part due to their simplicity and ease of decoding. 
Such schemes seek to separate users' transmissions across time (TDMA), frequency (FDMA), code (CDMA), or other orthogonal domains in such a way that avoids overlap between users~\cite{wittman1967multipleaccess}.
When OMA is used over the Gaussian multiple access channel (GMAC), it is possible for each user to achieve the finite block length (FBL) capacity~\cite{polyanskiy2010fbl} of its effective single-user AWGN channel by employing random Gaussian coding; however, this strategy has an overall complexity of $\Ocal\left(nK2^{n}\right)$, where $K$ denotes the number of users and $n$ is the blocklength of the single-user code. 
To meet the stringent latency requirements of wireless standards, it is common for users to employ more practical codes such as 5G-NR LDPC codes~\cite{richardson20185gldpc} instead.
This solution (OMA-LDPC) features a $\Ocal\left(nK\right)$ complexity that is linear in $K$ and $n$, but suffers from reduced FBL spectral efficiency. 

As an alternative to OMA, non-orthogonal multiple access (NOMA) has been proposed as a means to obtain a high spectral efficiency while maintaining reasonable complexity.
Furthermore, NOMA benefits from the ability to support more connected users than degrees of freedom in the system; thus, NOMA schemes are considered to be promising architectures for next-generation wireless systems~\cite{ding2017noma}.
While many NOMA techniques have been proposed in the literature, we restrict our attention to the class of NOMA schemes whose complexity is linear in $K$. 
An emblematic scheme from this class is interleave division multiple access (IDMA)~\cite{ping2006idma}. 

IDMA proceeds by first having users encode their messages using a low rate code amenable to soft decoding.
Then, users permute their codewords using unique interleavers, modulate their resultant bit sequences, and then simultaneously transmit their messages over the multiple access channel (MAC). 
The receiver then uses an iterative, turbo-style algorithm that passes extrinsic log likelihood ratios (LLR)s between a signal estimator and a bank of single-user decoders for recovering each user's transmitted signal.
This scheme has a complexity of $\Ocal\left(nK\right)$.
Note that, unlike OMA, the number of channel uses in this scheme does not need to scale as $n = n_1K$, where $n_1$ is the number of channel uses required to support a single user.
Though elegant, IDMA also suffers from reduced FBL spectral efficiency when compared to OMA with random Gaussian coding.

In this article, a novel NOMA scheme for the coordinated uplink channel called Multi-User SR-LDPC (MU-SR-LDPC) coding is presented.
For a fixed error rate, MU-SR-LDPC coding achieves a higher FBL spectral efficiency than both OMA-LDPC and IDMA with only a marginally higher complexity of $\Ocal\left(Kn\log\left(n\right)\right)$~\cite{ebert2024multiuser}, which is still linear in $K$.
Furthermore, it is observed that the FBL spectral efficiency of MU-SR-LDPC codes increases as the number of users increases.
In fact, for high enough $K$ and proper $n$, the spectral efficiency of MU-SR-LDPC coding exceeds that of OMA with optimal single-user random Gaussian coding. 
The proposed scheme consists of each user transmitting its own Sparse Regression LDPC (SR-LDPC)~\cite{ebert2023srldpc} codeword using a unique sensing matrix that is known both to the user and to the base station. 
The decoder then employs a low-complexity AMP-BP algorithm for jointly recovering the set of transmitted codewords using techniques from coded demixing~\cite{ebert2022codeddemixing}.

Before proceeding with our exposition of MU-SR-LDPC codes, we pause to briefly review pertinent prior work on SR-LDPC codes and coded demixing.

\subsection{Prior Work}

SR-LDPC codes are constructed by concatenating an inner sparse regression code (SPARC)~\cite{joseph2012least, joseph2013fast, venkataramanan2019sparse} with an outer non-binary LDPC (NB-LDPC) code~\cite{davey1998low, bennatan2006design}, where the field size of the LDPC code is carefully chosen to match the section size of the SPARC~\cite{ebert2023srldpc}.
SR-LDPC codewords are obtained by first encoding information bits using the NB-LDPC code, then mapping each coded symbol to a single SPARC section, then concatenating the sections together to form a block-sparse vector, and finally compressing the resultant vector using a suitable sensing matrix. 

SR-LDPC decoding consists of recovering a block-sparse vector from noisy measurements in the presence of an outer code. 
This may be accomplished using approximate message passing (AMP)~\cite{donoho2009messagepassing}, which is a low-complexity algorithm for sparse recovery~\cite{candes2008cs} that incorporates side information into the iterative recovery process~\cite{ma2019sideinfo}.
AMP is known to perform exceptionally well; in fact, it has been shown that SPARCs with AMP decoding can be capacity-achieving over the AWGN channel in certain situations~\cite{barbier2014amp, barbier2017approximate, rush2017capacity, rush2021capacity}. 
To account for the side information afforded by the outer code, SR-LDPC decoding is performed by an AMP algorithm whose denoiser runs belief propagation (BP) on the factor graph of the outer LDPC code during each AMP iteration~\cite{amalladinne2022ccs, ebert2023srldpc}. 

Interestingly, when the sensing matrix is a random Gaussian matrix, each SR-LDPC coded symbol is an independent Gaussian random variable, thus enabling SR-LDPC codes to exploit shaping gains. 
Under the aforementioned AMP-BP decoding algorithm, SR-LDPC codes have been shown to provide  competitive performance over AWGN channels of moderate block lengths~\cite{ebert2023srldpc}.

SR-LDPC codes are inspired by compressed sensing, which is a framework for recovering a single signal that is sparse with respect to a given domain.
A natural extension of compressed sensing is convex demixing, which seeks to recover multiple signals that are sparse with respect to distinct domains given a single set of superimposed measurements~\cite{mccoy2014sharp, zhou2017demixing}.
When the signals to be recovered each have an SR-LDPC-like structure, i.e., are block sparse with graphically linked sections, and the sparse domains exhibit sufficiently low cross-coherence, a modified AMP-BP algorithm may be used to jointly recover the constituent signals with low complexity~\cite{ebert2022codeddemixing}. 
This technique, known as coded demixing, forms the basis for the proposed MU-SR-LDPC recovery algorithm.

\subsection{Main Contributions}

This article introduces a novel NOMA scheme for the coordinated uplink channel called MU-SR-LDPC coding.
MU-SR-LDPC coding consists of each user transmitting its own SR-LDPC codeword using a unique sensing matrix known both to the user and the base station.
The receiver then employs coded demixing techniques to jointly recover the set of transmitted codewords. 
Through numerical simulations, it is shown that MU-SR-LDPC codes achieve a higher FBL spectral efficiency than both OMA-LDPC and IDMA with only an additional $\log(n)$ factor in complexity.
In fact, it is observed that MU-SR-LDPC codes may match the performance of maximum a posteriori (MAP) decoding in certain regimes without incurring exponential complexity. 
Finally, it is shown that for certain combinations of blocklength and number of users and for a fixed error rate, the spectral efficiency of MU-SR-LDPC coding exceeds that of OMA with single-user random Gaussian coding.
While this observation does not violate fundamental limits because OMA is known to be suboptimal for FBL, it is a highly desirable attribute for practical system, especially in view of the low-complexity decoding algorithm.

\subsection{Notation}

In the sequel, bold capital letters such as $\Am$ denote matrices and bold lowercase letters such as $\xv$ denote (column) vectors. 
Scalars are represented as lowercase letters as in $x$ and sets are denoted using caligraphic letters such as $\Scal$.
Two exceptions to this rule are $\Ncal\left(\mu, \sigma^2\right)$, which denotes a Gaussian distribution with mean $\mu$ and variance $\sigma^2$, and $[N]$, which denotes the set $\{0, 1, \ldots, N-1\}$. 
The expected value of function $f(X)$ with respect to random variable $X$ is denoted by $\mathbb{E}_X\left[f(X)\right]$, though the subscript is occasionally dropped when the source of the randomness is unambiguous from context. 

\section{System Model}
\label{section:system_model}

Consider a coordinated uplink multiple access channel consisting of $K$ active users that each wish to transmit coded messages over $n$ total channel uses (real degrees of freedom), where $n$ scales at most linearly with $K$. 
Each active user $k \in [K]$ encodes $B$ information bits $\wv_k \in \mathbb{F}_2^{B}$ into the codeword $\xv_k \in \mathbb{R}^{n}$, and then the active users synchronously transmit their codewords as coordinated by the base station. 
The signal $\yv$ received by the base station is thus given by
\begin{equation}
    \yv = \sum_{k \in [K]} \xv_k + \zv,
\end{equation}
where $\zv \sim \Ncal\left(0, \sigma^2\IDm\right)$ denotes additive white Gaussian noise (AWGN). 
For simplicity, we assume a uniform power allocation over channel uses so that $\mathbb{E}\left[\|\xv_k\|_2^2\right] = P$ and $\mathbb{E}\left[|\xv_k(i)|^2\right] = {P}/{n}$ for all $i \in [n]$.
Given $\yv$, the base station is tasked with recovering the set of transmitted messages $\{\wv_k : k \in [K]\}$.

Note that the assumption that sensing matrices are unique across users and known to both the users and the base station assumes a certain level of coordination between users and the base station. 
We assume that this coordination occurs, but do not consider how this coordination takes place.
We emphasize that this simplifying assumption is common in the literature; mechanisms already exists in cellular systems to facilitate coordination.
Next, we review the MU-SR-LDPC encoding and decoding procedures in detail~\cite{ebert2024multiuser}. 

\subsection{MU-SR-LDPC Encoding}

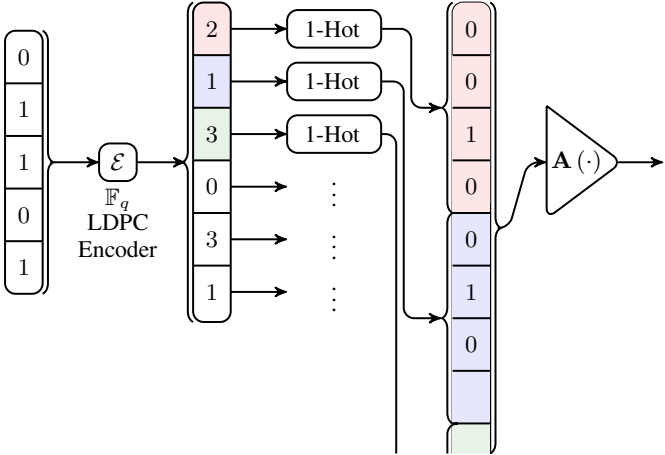
\begin{figure}[t]
    \centering
    \begin{tikzpicture}[
    font=\small,  >=stealth',
    line width = 0.75pt,
]

\definecolor{darkgreen}{RGB}{0,128,0}

\draw[rounded corners] (0, 0-0.375) rectangle (0.5, -3.5-0.375);
\draw[decorate, decoration={brace, amplitude=4pt}] (0.5, 0-0.375) -- (0.5, -3.5-0.375) {};
\foreach \j in {1, 2, 3, 4} {
    \draw[-] (0.0, -0.7*\j-0.375) to (0.5, -0.7*\j-0.375);
}
\node[] at (0.25, -1*0.7+0.35-0.375) {$0$};
\node[] at (0.25, -2*0.7+0.35-0.375) {$1$};
\node[] at (0.25, -3*0.7+0.35-0.375) {$1$};
\node[] at (0.25, -4*0.7+0.35-0.375) {$0$};
\node[] at (0.25, -5*0.7+0.35-0.375) {$1$};

\draw[rounded corners] (1.25, -1.5-0.375) rectangle (1.75, -2.0-0.375); 
\node[] at (1.5, -1.75-0.375) {$\mathcal{E}$};
\node[] at (1.5, -2.25-0.375) {$\mathbb{F}_q$};
\node[] at (1.5, -2.55-0.375) {LDPC};
\node[] at (1.5, -2.9-0.375) {Encoder};
\draw[->] (0.60, -1.75-0.375) to (1.25, -1.75-0.375);

\fill[red!10, rounded corners] (2.5, 0.0) rectangle (3.0, -0.7) {};
\fill[red!10] (2.5, -0.35) rectangle (3.0, -0.7) {};
\fill[blue!10] (2.5, -0.7) rectangle (3.0, -0.7*2) {};
\fill[darkgreen!10] (2.5, -0.7*2) rectangle (3.0, -0.7*3) {};

\draw[rounded corners] (2.5, 0) rectangle (3.0, -4.25);
\draw[decorate, decoration={brace, mirror, amplitude=4pt}] (2.5, 0) -- (2.5, -4.25) {};
\foreach \j in {1, 2, 3, 4, 5} {
    \draw[-] (2.5, 0-0.7*\j) to (3.0, 0-0.7*\j);
}

\node[] at (2.75, -1*0.7+0.35) {$2$};
\node[] at (2.75, -2*0.7+0.35) {$1$};
\node[] at (2.75, -3*0.7+0.35) {$3$};
\node[] at (2.75, -4*0.7+0.35) {$0$};
\node[] at (2.75, -5*0.7+0.35) {$3$};
\node[] at (2.75, -6*0.7+0.35) {$1$};
\draw[->] (1.75, -1.75-0.375) to (2.4, -1.75-0.375);

\foreach \j in {0, 1, 2} {
    \draw[rounded corners] (3.75, -0.7*\j-0.1) rectangle (5.0, -0.7*\j-0.5-0.1);
    \node[] at (4.35, -0.7*\j-0.1-0.25) {1-Hot};
    \draw[->] (3.0, -0.7*\j-0.1-0.25) to (3.75, -0.7*\j-0.1-0.25);
}
\foreach \j in {3, 4, 5} {
    \draw[->] (3.0, -0.7*\j-0.1-0.25) to (3.75, -0.7*\j-0.1-0.25);
    \node[] at (0.625+3.75, -0.7*\j-0.1-0.25) {$\vdots$}; 
}

\draw[rounded corners] (5.95, -6.0) to (5.95, 0) to (6.45, 0) to (6.45, -6.0);

\fill[red!10, rounded corners] (5.95, 0.0) rectangle (6.45, -0.7*4) {};
\fill[red!10] (5.95, -0.7) rectangle (6.45, -0.7*4) {};
\fill[blue!10] (5.95, -0.7*4) rectangle (6.45, -0.7*8) {};
\fill[darkgreen!10] (5.95, -0.7*8) rectangle (6.45, -6.0) {};

\foreach \j in {1, 2, 3, 4, 5, 6, 7, 8} {
    \draw[-] (5.95, -0.7*\j) to (6.45, -0.7*\j);
}

\node[] at (6.2, -0.7*1+0.35) {$0$};
\node[] at (6.2, -0.7*2+0.35) {$0$};
\node[] at (6.2, -0.7*3+0.35) {$1$};
\node[] at (6.2, -0.7*4+0.35) {$0$};
\node[] at (6.2, -0.7*5+0.35) {$0$};
\node[] at (6.2, -0.7*6+0.35) {$1$};
\node[] at (6.2, -0.7*7+0.35) {$0$};

\draw[decorate, decoration={brace, mirror, amplitude=4pt}] (5.95, 0) -- (5.95, -0.7*4) {};
\draw [rounded corners, ->] (5.0, -0.35) to (5.4, -0.35) to (5.4, -1.4) to (5.8, -0.7*2);

\draw[decorate, decoration={brace, mirror, amplitude=4pt}] (5.95, -0.7*4) -- (5.95, -0.7*8) {};
\draw[rounded corners, ->] (5.0, -1.3+0.25) -- (5.3, -1.3+0.25) -- (5.3, -0.7*6) -- (5.8, -0.7*6);

\draw[rounded corners] (5.95, -0.7*8) -- (5.88, -0.7*8) -- (5.88, -6.0) {};
\draw[rounded corners] (5.0, -1.75) -- (5.2, -1.75)  -- (5.2, -6.0) {};

\draw[decorate, decoration={brace, amplitude=4pt}] (6.45, 0) -- (6.45, -6.0) {};
\draw[rounded corners] (7.2, -1.0-0.375) -- (8.2, -1.75-0.375) -- (7.2, -2.5-0.375) -- (7.2, -1.0-0.375);
\node[] at (7.6, -1.75-0.375) {$\Am\left(\cdot\right)$};
\draw[rounded corners, ->] (6.55, -3.0) -- (6.70, -3.0) -- (6.95, -1.75-0.375) -- (7.2, -1.75-0.375);

\draw[->] (8.15, -1.75-0.375) -- (8.75, -1.75-0.375);

\end{tikzpicture}
    \vspace{-5mm}
    \caption{This figure depicts the encoding process for a single user in a MU-SR-LDPC system. Information messages are first encoded using a NB-LDPC code, then each coded symbol is transformed into a one-sparse index vector. The set of index vectors is then vertically concatenated and the resultant vector is pre-multiplied by the sensing matrix $\Am$ to obtain an SR-LDPC codeword.}
    \label{fig:encoding}
\end{figure}

Since the encoding operation is identical for all users, we temporarily drop the subscript $k$ throughout this section.
The MU-SR-LDPC encoding process begins with each user encoding its $B$-bit binary information message $\wv \in \mathbb{F}_2^B$ into a non-binary LDPC codeword 
\begin{equation}
    \vv = \left(v_0, v_1, \ldots, v_{L-1}\right) \in \mathbb{F}_q^L
\end{equation}
via well-studied operations~\cite{bennatan2006design}.
Here, $q = 2^p,~p\in\mathbb{N}$ is the order of the field and $L$ is the length of the code. 
Then, every symbol $v_{\ell} \in \vv$ is mapped to a $1$-sparse standard basis vector $\ev_{\phi(v_\ell)} \in \mathbb{R}^{q}$ in a manner akin to one-hot encoding, where $\phi(v): \mathbb{F}_q \mapsto [q]$ is an arbitrary bijection that maps $0 \in \mathbb{F}_q$ to $0 \in \mathbb{N}$ and $1 \in \mathbb{F}_q$ to $1 \in \mathbb{N}$. 
The $L$ resultant basis vectors are then vertically concatenated together to obtain
\begin{equation}
    \sv = \begin{bmatrix}
        \ev_{\phi(v_0)} \\
        \ev_{\phi(v_1)} \\
        \vdots \\
        \ev_{\phi(v_{L-1})}
    \end{bmatrix} \in \mathbb{R}^{qL}.
\end{equation}
Note that the vector $\sv$ has two significant structures: 
first, $\sv$ is block sparse, containing $L$ one-sparse sections each of length $q$, and second, each section of $\sv$ corresponds to exactly one coded symbol, and the coded symbols are connected together across sections via the outer LDPC code. 
These structures will be exploited by the decoding algorithm. 
After obtaining the vector $\sv$, the SR-LDPC codeword is obtained by pre-multiplying $\sv$ by a sensing matrix $\Am \in \mathbb{R}^{n \times qL}$ to obtain
\begin{equation}
    \xv = \Am\sv.
\end{equation}

In a MU-SR-LDPC system, each user $k \in [K]$ is assigned its own sensing matrix $\Am_k$ from a set of sensing matrices that exhibit low cross-coherence. 
In the sequel, it is assumed that every entry in each sensing matrix is generated i.i.d. according to $a_{i, j}\sim\Ncal\left(0, 1/n\right)$; this assumption will be important in the development of our AMP-based recovery algorithm~\cite{donoho2009messagepassing}.
However, in practice, it is common to use subsampled Hadamard matrices so that fast transform techniques may be used to reduce complexity. 
Fig.~\ref{fig:encoding} depicts the encoding process for a single user.

\subsection{MU-SR-LDPC Decoding}

We now turn our attention to the decoding operation performed at the base station.
Given the structure of the transmitted signals, the received signal may be expressed as
\begin{equation}
    \yv = \sum_{k \in [K]} \Am_k\sv_k + \zv.
\end{equation}
This equation may be written in matrix-vector form by horizontally stacking the $\{\Am_k : k \in [K]\}$ matrices to form
\begin{equation}
    \label{eq:phim}
    \Phim = \begin{bmatrix} \Am_0 & \hdots & \Am_{K-1}\end{bmatrix} \in \mathbb{R}^{n \times qLK}
\end{equation}
and similarly vertically stacking the $\{\sv_k: k \in [K]\}$ vectors to form
\begin{equation}
    \label{eq:sfrak}
    \sfrak = \begin{bmatrix}
        \sv_0 \\
        \vdots \\
        \sv_{K-1} \\
    \end{bmatrix} \in \mathbb{R}^{qLK}.
\end{equation}
Then, it follows that
\begin{equation}
    \label{eq:rx_signal_cs_form}
    \yv = \Phim \sfrak + \zv.
\end{equation}
Note that \eqref{eq:rx_signal_cs_form} assumes a canonical noisy compressed sensing form, albeit in an exceedingly large dimensional space. 
Thus, a viable MU-SR-LDPC decoding strategy is to first perform structured sparse recovery to recover $\sfrak$ from $\yv$, and subsequently to extract $\{\wv_k : k \in [K]\}$ from $\sfrak$. 
For the first part of the proposed decoding strategy, we wish to employ a low complexity sparse recovery algorithm that can efficiently handle the extreme dimensions of the problem while simultaneously exploiting the available side information to maximize performance. 

Approximate message passing (AMP) is a sparse recovery algorithm that is known to provide excellent performance at low complexity~\cite{donoho2009messagepassing}.
A crucial feature of AMP is its denoiser, which allows for side information to be incorporated into the iterative recovery process~\cite{ma2019sideinfo}. 
A specific version of AMP known as the AMP-BP algorithm uses a denoiser that runs belief propagation (BP) on the factor graph of the outer LDPC code within each AMP iteration~\cite{amalladinne2022ccs}.
This allows for soft information to be dynamically shared between the iterative sparse recovery algorithm and the NB-LDPC decoder and has been shown to provide a steep waterfall in error performance, a phenomenon not observed when inner and outer codes are decoded in disjoint succession~\cite{ebert2023srldpc}.
In the coded demixing literature, it is known that a modified AMP-BP algorithm may be used to efficiently recover multiple signals with SR-LDPC-like structure given a vector of superimposed measurements~\cite{ebert2022codeddemixing}.
In light of these facts, we propose to utilize a coded demixing algorithm with minimal modifications for MU-SR-LDPC decoding. 
We will now discuss the details of the envisioned AMP-BP algorithm.

\begin{figure*}[ht!]
    \centering
    \begin{tikzpicture}
  [
  font=\small, >=stealth', line width=1pt,
  check/.style={rectangle, minimum height=2.5mm, minimum width=2.5mm, draw=black},
  varnode/.style={circle, minimum size=2mm, draw=black},
  mmse/.style={rectangle, minimum height=7.5mm, minimum width=25mm, rounded corners, draw=black},
  quantity/.style={rectangle, minimum height=8mm, minimum width=8mm, rounded corners, draw=black},
  multiply/.style={trapezium, trapezium angle=75, draw=black, minimum width=10mm, minimum height=8mm, rounded corners}
  ]

\definecolor{darkgreen}{RGB}{0,128,0}

%
\node[quantity, label=Received Signal] (signal) at (-7,3.5) {$\mathbf{y}$};
\draw[*->, rounded corners] (-9.25, 4.5) -- (-9.25, 3.5) -- node[below] {Input} (signal);

%
\node[quantity] (residual) at (-5.5,2) {$\mathbf{z}^{(t)}$};
\node[circle, minimum width=8mm, draw=black] (sum) at (-7,2) {$\sum$}
  edge[<-] (signal)
  edge[->] (residual);
\node[quantity] (stack) at (-5,-2) {Stack};
\draw[*->, rounded corners] (stack) -- (-5,-4) -- node[above] {Output} (-3,-4);
\node[multiply,label=below:Amplitude] (Dmatrix) at (-7,-2) {$\Dm$}
  edge[<-] (stack);
\node[multiply] (Amatrix) at (-7,0.5) {$\Phim$}
  edge[->] node[right] {$-$} (sum)
  edge[<-] (Dmatrix);

\foreach \g/\xg/\yg in {3/3.5/2, 2/2.25/1,1/1/0} {

    \ifnum \g = 3
        \def\color{darkgreen};
    \fi
    \ifnum \g = 2
        \def\color{blue};
    \fi
    \ifnum \g = 1
        \def\color{red};
    \fi
    
    %
    \draw[draw=\color, densely dotted, rounded corners, fill=white] (-4.25+\xg,-0.25+\yg) rectangle (0.375+\xg,2.75+\yg);
    \node[multiply, shape border rotate=270, fill=\color!10] (dual\g) at (-3.5+\xg,2+\yg) {$\Am_\g^\intercal$};
    
    \draw[->, rounded corners, draw=\color] (residual.east) -- (-5.5+\xg,2+\yg) -- (dual\g);
    \node[circle, minimum width=8mm, draw=black] (sum\g) at (-2+\xg,2+\yg) {$\sum$}
      edge[<-, draw=\color] (dual\g);
    \node[quantity] (rv\g) at (-0.5+\xg,2+\yg) {$\rv_\g^{(t)}$}
      edge[<-, draw=\color] (sum\g)
      edge[->, draw=\color] (1+\xg,2+\yg);
    \node[multiply] (amplitude\g) at (-2+\xg,0.5+\yg) {$\Dm_\g$}
      edge[->, draw=\color] (sum\g);
    
    %
    \draw[draw=\color, densely dotted, rounded corners, fill=white] (1+\xg,-0.75+\yg) rectangle (4+\xg,2.75+\yg);
    \node[mmse] (mmse\g) at (2.5+\xg,-0.125+\yg){Dynamic PME};
    \draw[draw=black, rounded corners] (1.25+\xg,1+\yg) rectangle (3.75+\xg,2.5+\yg);
    \foreach \s in {1,2,3,4,5} {
      \node[varnode] (var\g-\s) at (1+0.5*\s+\xg,1.25+\yg) {}
        edge (mmse\g);
    }
    \foreach \c in {1,2,3} {
      \node[check] (check\g-\c) at (1+0.75*\c+\xg,2.25+\yg) {};
    }
    \draw (var\g-5) -- (check\g-3.south);
    \draw (var\g-4) -- (check\g-3.south);
    \draw (var\g-3) -- (check\g-3.south);
    \draw (var\g-2) -- (check\g-2.south);
    \draw (var\g-5) -- (check\g-2.south);
    \draw (var\g-4) -- (check\g-1.south);
    \draw (var\g-2) -- (check\g-1.south);
    \draw (var\g-1) -- (check\g-1.south);
    
    %
    \node[quantity] (state\g) at (-2+\xg,-3+\yg) {$\sv_\g^{(t)}$}
      edge[->, draw=\color] (amplitude\g);
    \draw[->,rounded corners, draw=\color] (state\g) -- (-4+\xg,-3+\yg) -- (stack.east);
    \node[quantity] (state\g-delay) at (2.5+\xg,-3+\yg) {Delay}
      edge[<-, draw=\color] (2.5+\xg,-0.75+\yg)
      edge[->, draw=\color] (state\g);
}


%
\draw[draw=black, densely dotted, rounded corners] (-10,-2.75) rectangle (-8.0, 0.25);
\node[draw=none,rotate=90] (onsagerlabel) at (-10.25, -1.25) {Onsager Term};
\node[quantity] (onsagerdelay) at (-5.5,-0.5) {Delay}
  edge(residual);
\node[quantity] (div) at (-9,-2) {$\frac{1}{n} \mathrm{div} (\cdot)$}
  edge[<-,dashed] (Dmatrix);
\node[circle, minimum width=8mm, draw=black] (times) at (-9,-0.5) {$\times$}
  edge[<-,dashed] (div)
  edge[<-,dashed] (onsagerdelay);
\draw[->,dashed,rounded corners] (times) -- (-9,2) -- (sum);

\end{tikzpicture}
    \caption{This figure provides a graphical representation of the proposed AMP-BP recovery algorithm which seeks to recover $\{\sv_k : k \in [K]\}$ from the vector of noisy observations $\yv$. The algorithm begins by computing a residual error enhanced with an Onsager correction term using all $K$ current state estimates. Then, each state estimate is refined based on the residual error and denoised by running a few rounds of BP on the factor graph of the outer LDPC code. The process then repeats itself until stopping conditions have been met. }
    \label{fig:mu_amp_bp}
\end{figure*}

The proposed AMP-BP algorithm is based on the standard AMP iterate~\cite{donoho2009messagepassing}, where
\begin{align}
    \zv^{(t)} &= \yv - \Phim \sfrak^{(t)} + \frac{\zv^{(t-1)}}{n} \operatorname{div} \hfrak^{(t-1)} \left( \rfrak^{(t-1)} \right) \label{eq:amp_residual} \\
    \rfrak^{(t)} &= \Phim^{\mathrm{T}} \zv^{(t)} + \sfrak^{(t)} \label{eq:effective_observation} \\
    \sfrak^{(t+1)} &= \hfrak^{(t)} \left( \rfrak^{(t)} \right) . \label{eq:state_update}
\end{align}
At a high level, the AMP algorithm begins by computing a \textit{residual error} under the current state estimate in~\eqref{eq:amp_residual}, enhanced with an Onsager correction term. 
Then, the residual error is used to refine the state estimate in~\eqref{eq:effective_observation} to create an \textit{effective observation}. 
Finally, the effective observation is passed through a \textit{denoiser} in~\eqref{eq:state_update} to obtain the next state estimate.
The initial conditions for the algorithm are $\rfrak^{(0)} = \sfrak^{(0)} = \zerov$, $\zv^{(0)} = \yv$, and every quantity for which $t < 0$ is equal to the zero vector.

A crucial part of the AMP algorithm is the denoiser $\hfrak^{(t)} (\rfrak^{(t)})$, which seeks to incorporate available side information regarding $\sfrak$ into the iterative recovery process.
Recall that $\sfrak$ is composed of $K$ vertically stacked SR-LDPC codewords, each of which is block sparse and whose blocks are connected together via an outer NB-LDPC code. 
Thus, the argument to the denoiser may be rewritten as
\begin{equation}
    \rfrak^{(t)} 
    = \begin{bmatrix} \Am_0^{\mathrm{T}} \zv^{(t)} + \sv_0^{(t)} \\ \vdots \\ \Am_{K-1}^{\mathrm{T}} \zv^{(t)} + \sv_{K-1}^{(t)} \end{bmatrix}
    = \begin{bmatrix} \rv_0^{(t)} \\ \vdots \\ \rv_{K-1}^{(t)} \end{bmatrix}
\end{equation}
where each effective observation has the form
\begin{equation}
    \rv_k^{(t)} = \begin{bmatrix} \rv_{k,0}^{(t)} \\ \vdots \\ \rv_{k,L-1}^{(t)} \end{bmatrix}.
\end{equation}
A natural candidate for the denoiser is the minimum mean square error (MMSE) denoiser $\mathbb{E}\left[ \sfrak | \rfrak^{(t)}\right]$; unfortunately, this denoiser is computationally intractable in part due to the presence of the $K$ outer LDPC codes.
A low complexity alternative to the MMSE denoiser would be a section-wise MMSE denoiser $\mathbb{E}\left[\sv_{k, \ell} | \rv^{(t)}_{k, \ell}\right]$, yet this denoiser completely fails to account for the outer code structure.
As discussed in~\cite{amalladinne2022ccs, ebert2023srldpc}, a pragmatic alternative that trades off performance and complexity is the BP denoiser, which computes an estimate for each $\sv_{k, \ell}$ based on the observations contained within the computation tree of the $k$th user's code, up to a certain depth.
Given that the details of the BP denoiser are treated in depth in~\cite{ebert2023srldpc}, we only summarize the operation of the denoiser, focusing on the extension of the BP denoiser to the multi-user case.

Since $\sfrak$ is composed of $K$ independent SR-LDPC codewords, we propose to employ a denoiser that is user-separable, so that 
\begin{equation*}
    \hfrak^{(t)} \left( \rfrak^{(t)} \right)
    = \begin{bmatrix} \etav^{(t)}_0 \left( \rv_0^{(t)} \right) \\ \vdots \\ \etav^{(t)}_{K-1} \left( \rv_{K-1}^{(t)} \right) \end{bmatrix},
\end{equation*}
where $\etav\left(\cdot\right)$ is the BP denoiser from~\cite{ebert2023srldpc}.
In this representation, each section $\rv_{k}^{(t)}$ in $\rfrak^{(t)}$ acts as a vector observation about the true value of $\sv_{k}$.
An enabling property of the AMP-BP algorithm is that asymptotically, 
\begin{equation}
    \rv_{k,\ell}^{(t)} \sim \sv_{k,\ell} + \tau^{(t)} \zetav^{(t)}_{k, \ell},
\end{equation}
where $\tau^{(t)}$ is a deterministic scalar and $\zetav_{k, \ell}^{(t)}\sim\Ncal\left(0, \IDm\right)$ is a vector comprised of i.i.d. standard normal entries.
This property holds under the presence of the Onsager correction term, the fact that $a_{i, j} \sim \Ncal\left(0, 1/n\right)$, and some smoothness conditions on the denoising function~\cite{berthier2020state}. 
In~\cite{ebert2023srldpc}, it is shown that these conditions are satisfied for the $\etav$ denoiser.
It is straightforward to extend the arguments from~\cite{ebert2023srldpc} to show that these conditions are similarly satisfied for the proposed $\hfrak$ denoiser.

Thus, by leveraging the relations
\begin{xalignat*}{2}
\rv_k^{(t)} &= \begin{bmatrix} \rv_{k,0}^{(t)} \\ \vdots \\ \rv_{k,L-1}^{(t)} \end{bmatrix} ; &
\sv_k^{(t)} &= \begin{bmatrix} \sv_{k,0}^{(t)} \\ \vdots \\ \sv_{k,L-1}^{(t)} \end{bmatrix},
\end{xalignat*}
we obtain a vector effective observation $\rv_{k, \ell}$ about $\sv_{k, \ell} \in \left\{ \ev_g : g \in \phi(\mathbb{F}_q) \right\}$.
This observation then induces a probability distribution on the $k$th user's $\ell$th coded symbol $\vv_{k, \ell}$, where
\begin{equation*}
    \begin{split}
        \boldsymbol{\alpha}_{k, \ell} (g)
        &= \Pr \left( \vv_{k, \ell} = g \Big| \rv_{k,\ell}^{(t)} \right)
        = \frac{e^{\frac{\rv_{k,\ell}^{(t)} \left( \phi(g) \right)}{\tau^2}}}
        {\sum_{h \in \mathbb{F}_q} e^{\frac{\rv_{k,\ell}^{(t)} \left( \phi(h) \right)}{\tau^2}}} .
    \end{split}
\end{equation*}
The collection of posterior probabilities $\{\alphav_{k, \ell} : \ell \in [L]\}$ may be used as local observations to initialize the factor graph of the $k$th user's outer LDPC code.
Then, estimates for $\{\vv_{k, \ell}: \ell \in [L]\}$ may be refined by successively running several rounds of BP on that same factor graph.
As the rest of the BP denoising process is identical across users, we again temporarily drop the subscript $k$ to lighten notation. 
Within the BP denoiser, the check to variable node messages are computed as 
\begin{equation} 
    \label{eq:check_to_var_msg}
    \boldsymbol{\mu}_{c_p \to v_{\ell}}
    = \left( \bigodot_{v_j \in N(c_p) \setminus v_{\ell}} \left( \boldsymbol{\mu}_{v_j \to c_p} \right)^{\times \omega_{j,p}^{-1}} \right)^{\times \left(- \omega_{\ell,p} \right)}
\end{equation}
and the variable to check messages are computed as
\begin{equation} 
    \label{eq:var_to_chk_msg}
    \boldsymbol{\mu}_{v_{\ell} \to c_p}
    = \frac{ \boldsymbol{\alpha}_{\ell} \circ \left( \operatorname*{\bigcirc}_{c_{\xi} \in N(v_{\ell}) \setminus c_p} \boldsymbol{\mu}_{c_{\xi} \to v_{\ell}} \right) }
    { \left\| \boldsymbol{\alpha}_{\ell} \circ \left( \operatorname*{\bigcirc}_{c_{\xi} \in N(v_{\ell}) \setminus c_p} \boldsymbol{\mu}_{c_{\xi} \to v_{\ell}} \right) \right\|_1 }.
\end{equation}
In the above message passing rules, $\odot$ denotes $\mathbb{F}_q$ convolution
\begin{equation}
    \left[ \boldsymbol{\mu} \odot \boldsymbol{\nu} \right]_{g}
    = \sum_{h \in \mathbb{F}_q} \mu_h \cdot \nu_{g - h}
    \qquad g \in \mathbb{F}_q,
\end{equation}
and $\bigcirc$ denotes the Hadamard product.
The special operator $\times g$ accounts for the impact of the field structure under multiplications within the belief vector, as explained in~\cite{bennatan2006design}; this action is well-established in the treatment of non-binary LDPC codes.
The state estimate at time $t+1$ is take to be, section-wise, the BP estimate 
\begin{equation} 
    \label{eq:out_msg}
    \sv_{\ell}^{(t+1)}
    = \frac{ \boldsymbol{\alpha}_{\ell} \circ \left( \operatorname*{\bigcirc}_{c_{p} \in N(v_{\ell}) } \boldsymbol{\mu}_{c_{p} \to v_{\ell}} \right) }
    { \left\| \boldsymbol{\alpha}_{\ell} \circ \left( \operatorname*{\bigcirc}_{c_{p} \in N(v_{\ell}) } \boldsymbol{\mu}_{c_{p} \to v_{\ell}} \right) \right\|_1 }.
\end{equation}
After vertically stacking the $L$ section estimates associated with each user, the $K$ users' state estimates are then stacked vertically to obtain $\sfrak^{(t+1)}$ and the AMP iterative process continues.
Fig.~\ref{fig:mu_amp_bp} graphically depicts the decoding process. 

Assuming that the number of BP iterations performed per AMP iteration is less than the girth of the factor graph~\cite{ebert2023srldpc}, the divergence of the denoiser may be expressed as
\begin{equation}
    \begin{split}
        \operatorname{div}\hfrak^{(t)} \left( \rfrak^{(t)} \right)
        &= \sum_{k \in [K]} \operatorname{div} \etav^{(t)}_{k} \left( \rv_k^{(t)} \right) \\
        &\hspace{-15mm}= \sum_{k \in [K]} \frac{1}{\tau_t^2} \left( \left\| \etav_k^{(t)} \left( \rv^{(t)}_k \right) \right\|_1 - \left\| \etav_k^{(t)} \left( \rv_k^{(t)} \right) \right\|_2^2 \right) .
    \end{split}
\end{equation}
While the exact form of the Onsager term may be theoretically inconsequential, the fact that it is separable across users enables a significant parallelization of the proposed algorithm. 

Specifically, adopting a user-centric view of the decoding process, we may rewrite equations~\eqref{eq:amp_residual}-\eqref{eq:state_update} as
\begin{align}
    \hat{\yv}_k^{(t)} = \Am_k \sv_k^{(t)} &- \frac{1}{n} \zv^{(t-1)} \operatorname{div} \etav_{k}^{(t-1)} \left( \rv_k^{(t-1)} \right) \label{eq:EstimatedContributions} \\
    \zv^{(t)} &= \textstyle \yv - \sum_{k \in [K]} \hat{\yv}_k^{(t)} \label{eq:AMP-Residual-Plus} \\
    \rv_{k}^{(t)} &= \Am_k^{\mathrm{T}} \zv^{(t)} + \sv_k^{(t)} \label{eq:EffectiveObservation-Plus} \\
    \sv_{k}^{(t+1)} &= \etav_{k}^{(t)} \left( \rv_{k}^{(t)} \right) \label{eq:AMP-Denoising-Plus} .
\end{align}
In the above formulation, the set $\{\hat{\yv}_k^{(t)} : k \in [K]\}$ in~\eqref{eq:EstimatedContributions} may be computed in parallel, after which the collection of estimates are brought together to compute the joint residual error in~\eqref{eq:AMP-Residual-Plus}.
In this process, each user seeks to estimate its contribution to the received signal $\yv$ based on the residual error, the parity constraints of its own LDPC code, and its previous state estimate.
Notably, the computation of the joint residual requires the sharing of only one vector $\hat{\yv}_k^{(t)}$ per user.

Assuming that the sensing matrices $\{\Am_k : k \in [K]\}$ are each generated by randomly subsampling the rows of a Hadamard matrix, the computational complexity of matrix-vector multiplication is $\Ocal\left(N\log\left(N\right)\right)$, where $N = qL$. 
Furthermore, since the complexity of computing the Onsager correction term and performing BP denoising is only $\Ocal\left(N\right)$, it follows that the per-user complexity of a single state update is $\Ocal\left(N\log\left(N\right)\right)$. 
Since the complexity of computing the joint residual is $\Ocal\left(NK\right)$, it follows that the total per-AMP-iteration complexity of the proposed AMP-BP algorithm is $\Ocal\left(KN\log\left(N\right)\right)$, or equivalently, $\Ocal\left(nK\log\left(n\right)\right)$.
Thus, we conclude that the complexity of MU-SR-LDPC coding is only a $\log(n)$ factor higher than the considered benchmarks. 

\section{Performance Analysis}
\label{section:performance_analysis}

In this section, we consider the performance of MU-SR-LDPC codes over the AWGN channel.
As benchmarks, we compare MU-SR-LDPC coding to a generic orthogonal multiple access (OMA) scheme such as TDMA, FDMA, or CDMA. 
In the envisioned OMA scheme, each of the $K$ users encodes their messages using a $(n/K, B)$ single-user error correcting code, after which the coded bits are modulated and then assigned orthogonal time/frequency/code resources. 
After the users' codewords are synchronously transmitted through the GMAC, the decoder simply separates the $K$ users' transmissions and then decodes each codeword independently. 
We consider two choices of single user error correcting codes: 5G-NR LDPC codes and SR-LDPC codes.

We also compare MU-SR-LDPC coding to a low-complexity NOMA scheme called interleave division multiple access (IDMA)~\cite{ping2006idma}.
In IDMA, each user first encodes its information message using a low rate code amenable to soft decoding. 
Then, each user permutes its codeword using a unique interleaver, modulates the resultant bit sequence, and transmits its coded message over the GMAC. 
At the receiver, the decoder employs an iterative turbo-style algorithm that passes extrinsic LLRs between a signal estimator and individual single-user decoders.

We begin our study by considering a scenario with two users, each wishing to transmit $B = 584$ information bits over $n = 1460$ channel uses. 
We define the spectral efficiency, or sum rate, to be
\begin{equation}
    R_{\mathrm{sum}} = \frac{BK}{n};
\end{equation}
thus, it follows that $R_{\mathrm{sum}} = 0.8$ in this setting.

\begin{figure}
    \centering
    \begin{tikzpicture}

\definecolor{customred}{rgb}{0.63529,0.07843,0.18431} 
\definecolor{customblue}{rgb}{0.00000,0.44706,0.74118} 
\definecolor{customgreen}{rgb}{0.00000,0.49804,0.00000} 

\begin{semilogyaxis}[
    font=\small,
    width=7cm,
    height=5cm,
    scale only axis,
    every outer x axis line/.append style={white!15!black},
    every x tick label/.append style={font=\color{white!15!black}},
    xmin=0,
    xmax=5,
    xtick = {0, 1, 2, 3, 4, 5, 6, 7, 8},
    xlabel={$E_b/N_0$~(dB)},
    xmajorgrids,
    every outer y axis line/.append style={white!15!black},
    every y tick label/.append style={font=\color{white!15!black}},
    ymin=0.001,
    ymax=0.5,
    ytick = {0.0001, 0.001, 0.01, 0.1, 1},
    ylabel={BER},
    ymajorgrids,
    yminorgrids,
    legend style={at={(1,1)},anchor=north east, draw=black,fill=white,legend cell align=left}
]

\addplot [
    color=customgreen,
    solid,
    line width=2.0pt,
    mark size=2.5pt,
    mark=square,
    mark options={solid}
]
table[row sep=crcr]{
    0.0 0.25790 \\
    0.25 0.24870 \\
    0.5 0.24058  \\
    0.75 0.22879  \\
    1.0 0.21597  \\
    1.25 0.20071  \\
    1.5 0.17921  \\
    1.75 0.15249  \\
    2.0 0.11667 \\
    2.25 0.078586 \\
    2.5 0.038760  \\
    2.75 0.016225  \\
    3.0 0.0050717  \\
    3.25 0.0012190 \\
    3.5 0.00020510 \\
};
\addlegendentry{OMA  5G-LDPC};

\addplot [
    color=blue,
    solid,
    line width=2.0pt,
    mark size=2.5pt,
    mark=star,
    mark options={solid}
]
table[row sep=crcr]{
    0.0 0.2715 \\
    0.25 0.264 \\
    0.5 0.2454 \\
    0.75 0.2365 \\
    1 0.212 \\
    1.25 0.1875 \\
    1.5 0.166 \\
    1.75 0.1298 \\
    2 0.089219 \\
    2.25 0.0518 \\
    2.5 0.0275 \\
    2.75 0.0112 \\
    3.0 0.00335 \\
    3.25 0.000919 \\
};
\addlegendentry{OMA SR-LDPC};

\addplot [
    color=black,
    solid,
    line width=2.0pt,
    mark size=2.5pt,
    mark=triangle,
    mark options={solid}
]
table[row sep=crcr]{
    0.0 0.22601027 \\
    0.25 0.21993151 \\
    0.5 0.21401541 \\
    0.75 0.20779109499999998 \\
    1.0 0.198510275 \\
    1.25 0.189452055 \\
    1.5 0.17265411 \\
    1.75 0.16272111 \\
    2.0 0.147498095 \\
    2.25 0.09415681000000001 \\
    2.5 0.055238225 \\
    2.75 0.02362254 \\
    3.0 0.007134705 \\
    3.25 0.00143064 \\ 
    3.5 0.00034017499999999996 \\ 
};
\addlegendentry{IDMA};

\addplot [
    color=cyan,
    solid,
    line width=2.0pt,
    mark size=2.5pt,
    mark=triangle,
    mark options={solid}
]
table[row sep=crcr]{
    0.0 0.224349315 \\
    0.25 0.21470034500000001 \\
    0.5 0.20829623 \\
    0.75 0.199871575 \\
    1.0 0.18940925 \\
    1.25 0.17321061999999998 \\
    1.5 0.153629495 \\
    1.75 0.12988517500000002 \\
    2.0 0.077528905 \\
    2.25 0.03823438 \\
    2.5 0.008742610000000001 \\
    2.75 0.00171143 \\
    3.0 0.00026808 \\ 
};
\addlegendentry{MAP-IDMA};

\addplot [
    color=red,
    solid,
    line width=2.0pt,
    mark size=2.5pt,
    mark=diamond,
    mark options={solid}
]
table[row sep=crcr]{
    0.0 0.27 \\
    0.25 0.26 \\
    0.5 0.249 \\
    0.75 0.227 \\
    1.0 0.218 \\
    1.25 0.19 \\
    1.5 0.16 \\
    1.75 0.13 \\
    2.0 0.07 \\
    2.25 0.037
    2.5 0.01 \\
    2.75 0.002 \\
    3.0 0.0003 \\
};
\addlegendentry{MU-SR-LDPC};

\end{semilogyaxis}
\end{tikzpicture}
    \vspace{-5mm}
    \caption{This figure compares the BER performance of four candidate multiple access schemes: OMA with 5G-NR LDPC codes, OMA with SR-LDPC codes, IDMA with LDPC codes, and MU-SR-LDPC codes. In each case, there are $K = 2$ active users who each wish to transmit $B = 584$ information bits over $n = 1460$ channel uses (r.d.o.f.). This MU-SR-LDPC code requires a $0.3-0.5$dB lower SNR to achieve a target BER of $10^{-3}$ than its competitors.}
    \label{fig:single_cell_ber_comparison}
\end{figure}
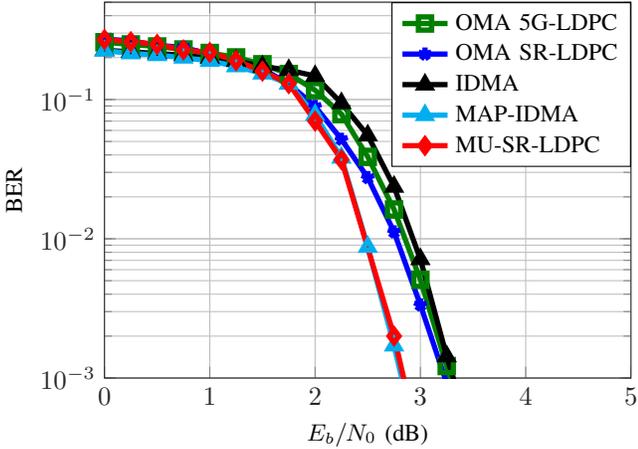

In the OMA-5G LDPC benchmark, we assume a TDMA system in which each user employs a $(730, 584)$ 5G LDPC code. 
The OMA SR-LDPC benchmark makes a similar TDMA assumption, only replacing the LDPC code with a $(730, 584)$ SR-LDPC code that uses a $(76, 73)$ outer NB-LDPC code over $\mathbb{F}_{256}$. 
The Tanner graph for the latter high-rate outer code is straightforward; it was generated using progressive edge growth (PEG)~\cite{hu2005peg} and then each edge was assigned a weight drawn uniformly at random from $\mathbb{F}_{256}\setminus\{0\}$. 
In the IDMA benchmark, each user employs a $(1460, 584)$ LDPC code and uses BPSK modulation.
Finally, the MU-SR-LDPC code uses the same $(76, 73)$ NB-LDPC code as the OMA SR-LDPC benchmark. 
Fig.~\ref{fig:single_cell_ber_comparison} plots the bit error rate (BER) as a function of $E_b/N_0$ for each of these schemes. 
The results are clear: this MU-SR-LDPC code requires a $0.3-0.5$dB lower SNR than the baseline schemes to achieve a target BER of $10^{-3}$ in this setting. 

Fig.~\ref{fig:single_cell_ber_comparison} also compares MU-SR-LDPC coding to IDMA under maximum a posteriori (MAP) decoding, whose complexity is exponential in $K$. 
In this regime, this MU-SR-LDPC code matches the performance of IDMA with MAP decoding at moderate to high SNRs.
This suggests that MU-SR-LDPC coding is a viable strategy to obtain MAP-level performance at sub-exponential complexity in certain regimes and access schemes. 

\begin{figure} [t!]
    \centering
    \begin{tikzpicture}

\definecolor{customred}{rgb}{0.63529,0.07843,0.18431} 
\definecolor{customblue}{rgb}{0.00000,0.44706,0.74118} 
\definecolor{customgreen}{rgb}{0.00000,0.49804,0.00000} 

\begin{semilogyaxis}[
    font=\small,
    width=7cm,
    height=5cm,
    scale only axis,
    every outer x axis line/.append style={white!15!black},
    every x tick label/.append style={font=\color{white!15!black}},
    xmin=0.72,
    xmax=1.06,
    xtick = {0.6, 0.64, ..., 1.06},
    xlabel={$R_{\mathrm{sum}}$},
    xmajorgrids,
    every outer y axis line/.append style={white!15!black},
    every y tick label/.append style={font=\color{white!15!black}},
    ymin=0.0001,
    ymax=0.3,
    ytick = {0.0001, 0.001, 0.01, 0.1, 1.0},
    ylabel={BER},
    ymajorgrids,
    yminorgrids,
    legend columns=3,
    legend style={at={(0,0)},anchor=south west, draw=black,fill=white,legend cell align=left}
]

\addplot [
    color=orange,
    solid,
    line width=2.0pt,
    mark size=2.5pt,
    mark=diamond,
    mark options={solid}
]
table[row sep=crcr]{
    0.76 0.0011945 \\
    0.78 0.0017678 \\
    0.80 0.00327552 \\
    0.82 0.00558993 \\
    0.84 0.00819545 \\
    0.86 0.01403303 \\
    0.88 0.02377755 \\
    0.90 0.03274615 \\
    0.92 0.04644376 \\
    0.94 0.0623388 \\
    0.96 0.08271499 \\
    0.98 0.10532255 \\
    1.00 0.12312875 \\
    1.02 0.13892605 \\
    1.04 0.1702492 \\
    1.06 0.19568743 \\
};
\addlegendentry{$K = 1$};

\addplot [
    color=customgreen,
    solid,
    line width=2.0pt,
    mark size=2.5pt,
    mark=square,
    mark options={solid}
]
table[row sep=crcr]{
    0.84 0.001113375 \\
    0.86 0.0031113 \\
    0.88 0.0060271249999999995 \\
    0.90 0.01194792 \\
    0.92 0.022913125 \\
    0.94 0.038125275 \\
    0.96 0.05699058 \\
    0.98 0.08994847 \\
    1.00 0.10779853 \\
    1.02 0.148500325 \\
    1.04 0.16324114499999998 \\
    1.06 0.193645825 \\
};
\addlegendentry{$K = 2$};

\addplot [
    color=customblue,
    solid,
    line width=2.0pt,
    mark size=3.5pt,
    mark=asterisk,
    mark options={solid}
]
table[row sep=crcr]{
    0.88 0.0007324425 \\
    0.90 0.00229928 \\
    0.92 0.006512395 \\
    0.94 0.01674834 \\
    0.96 0.034953155 \\
    0.98 0.053700532499999995 \\
    1.00 0.1035140175 \\
    1.02 0.13348099000000002 \\
    1.04 0.1621659575 \\
    1.06 0.207042355 \\
};
\addlegendentry{$K = 4$};

\addplot [
    color=customred,
    solid,
    line width=2.0pt,
    mark size=2.5pt,
    mark=triangle,
    mark options={solid}
]
table[row sep=crcr]{
    0.92 0.0009512450000000001 \\
    0.94 0.00368026375 \\
    0.96 0.013962382499999999 \\
    0.98 0.042223131250000004 \\
    1.00 0.08213485499999999 \\
    1.02 0.13703216625 \\
    1.04 0.18077521875 \\
    1.06 0.1965896125 \\
};
\addlegendentry{$K = 8$};

\addplot [
    color=violet,
    solid,
    line width=2.0pt,
    mark size=2.5pt,
    mark=pentagon,
    mark options={solid}
]
table[row sep=crcr]{
    0.94 0.0004289625 \\
    0.96 0.005024113125 \\
    0.98 0.02256342625 \\
    1.00 0.08154843437500002 \\
    1.02 0.14196043062500002 \\
    1.04 0.17621187875 \\
    1.06 0.2134032525 \\
};
\addlegendentry{$K = 16$};

\addplot [
    color=teal,
    solid,
    line width=2.0pt,
    mark size=2.5pt,
    mark=Mercedes star,
    mark options={solid}
]
table[row sep=crcr]{
    0.76 0.0014 \\
    0.8 0.0067 \\
    0.84 0.028 \\
    0.877 0.055 \\
    0.916 0.116 \\
    0.959 0.16 \\
    0.998 0.17 \\
};
\addlegendentry{IDMA};

\addplot [
    color=black,
    dashed,
    line width=2.0pt
]
table[row sep=crcr]{
    0.89728616 0.000001 \\
    0.89728616 1.0 \\
};
\addlegendentry{$C_{\mathrm{OMA}}^{\mathrm{FBL}}$};

\addplot [
    color=black,
    solid,
    line width=2.0pt
]
table[row sep=crcr]{
    1.006188050069485 0.000001 \\
    1.006188050069485 1.0 \\
};
\addlegendentry{$C_{\mathrm{OMA}}$};

\end{semilogyaxis}
\end{tikzpicture}
    \vspace{-5mm}
    \caption{This figure plots the BER as a function of $R_{\mathrm{sum}}$ for the MU-SR-LDPC scheme under consideration at $E_b/N_0 = 3$dB. 
    This figure shows that MU-SR-LDPC coding achieves a higher spectral efficiency than its competitors.
    Furthermore, the spectral efficiency of MU-SR-LDPC coding appears to increase with $K$. }
    \label{fig:single_cell_spectral_efficiency}
\end{figure}
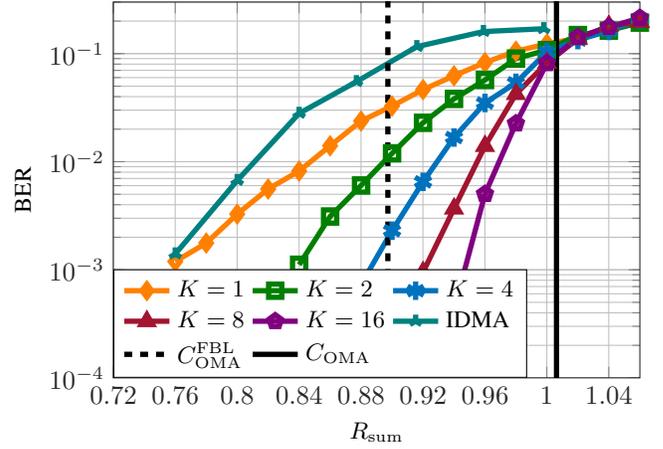

The results of Fig.~\ref{fig:single_cell_ber_comparison} suggest that this MU-SR-LDPC code may achieve a higher spectral efficiency than the alternatives at a target BER of $10^{-3}$; thus, we turn our attention to studying the spectral efficiency of these schemes for larger systems.
Throughout the study, we continue operating under our implicit assumption that all users employ the same rate for their uplink transmissions.

When an orthogonal multiple access scheme is employed, the number of channel uses $n$ must scale as $n = n_1K$; thus, the sum rate is equal to the per-user rate, or 
\begin{equation}
    R_{\mathrm{sum}} = \frac{BK}{n} = \frac{BK}{n_1K} = \frac{B}{n_1}.
\end{equation}
This is not necessarily the case in NOMA schemes because the number of users served may exceed the degrees of freedom in the system.
Indeed, the number of channel uses employed by a MU-SR-LDPC code is determined by the height of the matrix $\Phim$, which may be easily adjusted. 
Similarly, the number of channel uses employed in IDMA is determined by the block length of the LDPC code employed by each active user.

Fig.~\ref{fig:single_cell_spectral_efficiency} plots the BER as a function of $R_{\mathrm{sum}}$ at a fixed $E_b/N_0 = 3$dB and fixed $B = 584$ for a variety of scenarios. 
As the number of information bits is fixed, the rate is scaled by appropriately adjusting the blocklength.
First, the performance of IDMA, OMA SR-LDPC (labelled as $K = 1$), and MU-SR-LDPC coding (labelled as $K = 2$) for the two-user scenario is investigated. 
These empirical curves are plotted alongside the asymptotic capacity $C_{\mathrm{OMA}}$ of the single-user AWGN channel seen by each user within the OMA scheme as well as the approximate FBL capacity $C_{\mathrm{OMA}}^{\mathrm{FBL}}$ for that same channel at an error rate of $\epsilon = 10^{-3}$ and a blocklength of $n_1 = 768$.
From this comparison, we conclude that MU-SR-LDPC coding achieves about a $10\%$ higher spectral efficiency than both competing schemes.

Fig.~\ref{fig:single_cell_spectral_efficiency} also plots the spectral efficiency of MU-SR-LDPC coding when $K = 1, 2, 4, 8,$ and $16$. 
From this comparison, it is clear that the spectral efficiency of MU-SR-LDPC coding increases with $K$. 
In fact, when $K \in \{8, 16\}$, the spectral efficiency of MU-SR-LDPC coding is greater than the $C_{\mathrm{OMA}}^{\mathrm{FBL}}$ at a BER of $10^{-3}$, thus implying that MU-SR-LDPC coding outperforms all OMA schemes in this regime, regardless of complexity.

A possible explanation for the high spectral efficiency of MU-SR-LDPC coding is as follows. 
Whereas the block length of any one user's code remains constant in an OMA scheme, the block length of each user's SR-LDPC code increases with $K$ in a MU-SR-LDPC system.
It appears that the increased block length associated with higher $K$ translates into a higher coding gain, thus enabling a increase in spectral efficiency. 
A similar observation was made in~\cite{liang2020compressed}. 
This observation is consistent with the well-established fact that OMA schemes are not optimal for finite block lengths.

\section{Conclusion}
\label{section:conclusion}

In this article, a novel non-orthogonal multiple access scheme for the coordinated multi-user uplink channel is introduced called multi-user SR-LDPC (MU-SR-LDPC) coding.
MU-SR-LDPC coding consists of each user transmitting its own  SR-LDPC codeword using a sensing matrix known both to that user and the base station. 
The receiver then employs a highly parallelizable AMP-BP algorithm to jointly recover the set of transmitted messages.

MU-SR-LDPC codes are shown to achieve a higher spectral efficiency than comparable orthogonal and non-orthogonal multiple access schemes with only an additional $\log(n)$ factor in order-wise complexity.
Furthermore, MU-SR-LDPC codes are shown to match the performance of maximum a posteriori (MAP) decoding in certain regimes.
Finally, for certain regimes of blocklength $n$ and number of users $K$, MU-SR-LDPC codes are shown to achieve a higher spectral efficiency than all possible OMA schemes.
 
It is possible that the performance of MU-SR-LDPC codes may be further improved through a careful optimization of the code structure. 
For example, the outer LDPC code employed in this article was simply generated using PEG~\cite{hu2005peg}; thus, a thorough optimization of the outer LDPC code structure using convex optimization or density evolution techniques may improve system performance. 
Furthermore, it is well-known that the performance of SPARCs is significantly improved when a non-uniform power allocation is employed~\cite{venkataramanan2019sparse}.
Yet, this article makes no attempt to optimize the power allocation; thus, this remains a promising avenue for future research.  

\bibliographystyle{IEEEbib}
\bibliography{IEEEabrv, journal}

\end{document}